# High Speed Mid-Wave Infrared Uni-traveling Carrier Photodetector

Jian Huang, Zhiyang Xie, Yaojiang Chen, John E. Bowers, Baile Chen

*Abstract*—A mid-wave infrared (MWIR) frequency comb is expected to dramatically improve the precision and sensitivity of molecular spectroscopy. For high resolution applications, a high speed MWIR photodetector is one of the key components, however, commercially available high speed MWIR photodetectors only have sub-GHz bandwidth currently. In this paper, we demonstrate, for the first time to our knowledge, a high speed mid-wave infrared (MWIR) uni-traveling carrier photodetector based on InAs/GaSb type-II superlattice (T2SL) at room temperature. The device exhibits a cutoff wavelength of 5.6 μm and a 3dB bandwidth of 6.58 GHz for a 20 μm diameter device at 300 K. These promising results show that the device has potential to be utilized in high speed applications such as frequency comb spectroscopy, free space communication and others. The limitations on the high frequency performance of the photodetectors are also discussed.

## I. INTRODUCTION

MID-wave infrared (MWIR) spectroscopy is a powerful tool to diagnose composition of the chemicals due to the strong vibrational transitions in the mid-infrared domain. Frequency combs in the mid-wave infrared domain bring a new set of tools for precision spectroscopy, which could be used to precisely investigate the changes of composition of a molecular sample over a large dynamic range [1-6]. A high speed photodetector (PD) operated in the MWIR range is one of the key components for the MWIR frequency comb applications, which could significantly improve the resolution of the system. In addition, high speed MWIR PDs are also needed in other emerging application areas such as free space communication [7, 8] and mid-infrared light detection and ranging (LIDAR) systems. Therefore, the development of a high speed MWIR PD is essential in order to meet the increasing demands in these emerging areas. HgCdTe PDs have made good progress in achieving high quantum efficiency and high sensitivity photodetectors. However, they are often operated at low temperatures and have a relatively low bandwidth, which is not suitable for high speed application [9, 10]. Moreover, HgCdTe photodetectors are incompatible with the global initiative to eventually phase out the use of mercury. In recent years, high speed InSb photodetectors [11], inter-band cascade infrared photodetectors (ICIPs) [12, 13], quantum cascade photodetectors (QCDs) [14], and quantum well infrared photodetectors (QWIPs) [15-17] have been demonstrated. Ibrhim et al. reported an InSb p-i-n photodetector grown on GaAs with an electrical 3dB bandwidth of 8.5 GHz at room temperature [11]. Lotfi et al. showed a 1.3 GHz 3dB bandwidth ICIP using InAs/GaSb /AlSb/InSb based type-II superlattices (T2SLs) as active layer and cutoff around 4.2 μm at 300 K [12]. Chen et al. demonstrated a 2.4 GHz bandwidth ICIP employing an InAs/GaAsSb T2SLs absorption layer with a cutoff wavelength of 5.6 μm at 300K [13]. The QCD structures have also achieved a 3dB bandwidth of 5 GHz with peak response wavelength of 4.5 μm [14]. In addition, InGaAs/AlGaAs and GaAs/AlGaAs based QWIPs with RF response at 5 GHz and peak wavelength of 4.9 μm and 9 μm respectively have also been demonstrated [16, 17].

In this work, we report the InAs/GaSb T2SL based uni-traveling carrier (UTC) photodiodes for MWIR high speed application. The UTC photodiode structure can overcome the slow transport of optical generated hole by using p-type absorber layer, where the optical generated holes can quickly be collected within the dielectric relaxation time, and only the electron transit time dominates the total transit time of the device, thus it is expected to have higher speed than traditional PIN photodetectors [18-24]. A 20 μm-diameter device with 3dB bandwidth of 6.58 GHz and cutoff wavelength about 5.6 μm at 300K is demonstrated. To the best of our knowledge, this is the first UTC photodetector demonstrated in the MWIR band. The limiting factors of the 3dB bandwidth are also discussed by analyzing the small signal equivalent circuit of the devices.

This work was supported by National Key Research and Development Program of China (2018YFB2201000); National Natural Science Foundation of China (61975121); Strategic Priority Research Program of Chinese Academy of Sciences (Grant No. XDA18010000); Shanghai Sailing Program (17YF1429300) and ShanghaiTech University startup funding (F-0203-16-002). (Corresponding authors: Baile Chen.)

Jian Huang, Zhiyang Xie, Yaojiang Chen and Baile Chen are with the School of Information Science and Technology, ShanghaiTech University,Shanghai 201210, Jian Huang is also with the Shanghai Institute of Microsystem and Information Technology, Chinese Academy of Sciences, Shanghai 200050, China, and also with University of Chinese Academy of Sciences, Beijing 100049, China (corresponding author to provide e-mail: chenbl@shanghaitech.edu.cn).

John E. Bowers is with the Department of Electrical and Computer Engineering, University of California Santa Barbara, Santa Barbara 93106, USA.

## II. Device Growth and Fabrication

The epitaxial structure and band diagram of the designed device are shown in Fig. 1(a) and (b), respectively. The sample was grown on a n type GaSb substrate by molecular beam epitaxy system (MBE), and the epitaxial growth started with a 200 nm thick GaSb buffer layer. After that, a 100 nm n type (1×10¹⁸ cm⁻³) InAs/AlSb SL bottom contact layer and 100 nm n type (1×10¹⁸ cm⁻³) InAs/AlSb SL layer were grown followed by a 430 nm thick un-intentionally doped (u.i.d.) InAs/AlSb SL drift layer. Then, a 50 period InAs/GaSb SL grading p-doped absorption layer with doping concentration grading from 2×10¹⁸ cm⁻³ to uid was grown in order to have a self-induced electric field to facilitate the electron transport. That is followed by 10 periods of p type (2×10¹⁸ cm⁻³) AlAs$_{0.08}$Sb$_{0.92}$/GaSb SL as an electron blocking layer to prevent electrons in the absorption layer diffusing to p type (5×10¹⁸ cm⁻³) GaSb top contact layer.

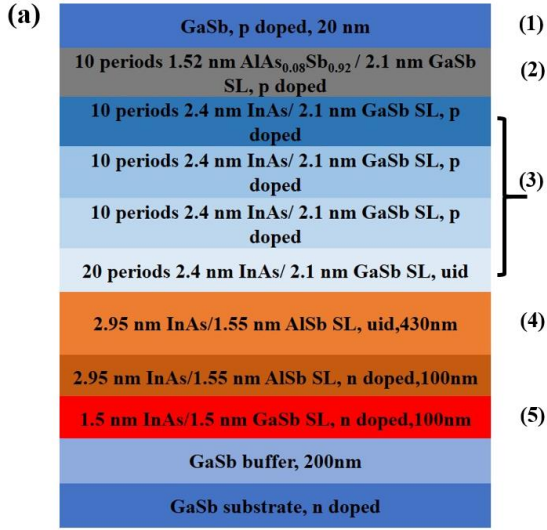

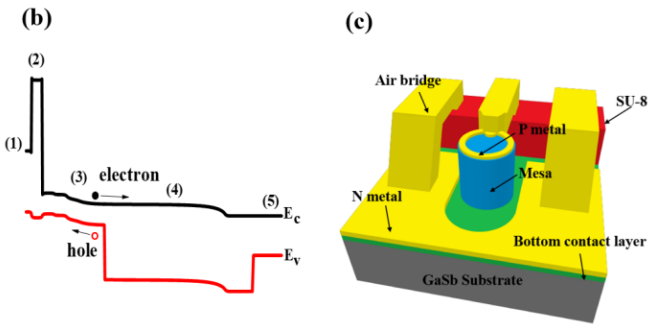

Fig. 1. (a) Epitaxial structure of the UTC PD. (b) Schematic band diagram of the UTC PD under zero bias. These different layers are: (1) top contact layer, (2) block layer, (3) graded absorption layer, (4) drift layer, (5) bottom contact layer. E$_c$ and E$_v$ represent conduction and valence band edge, respectively. (c) Schematic diagram of a fabricated device.

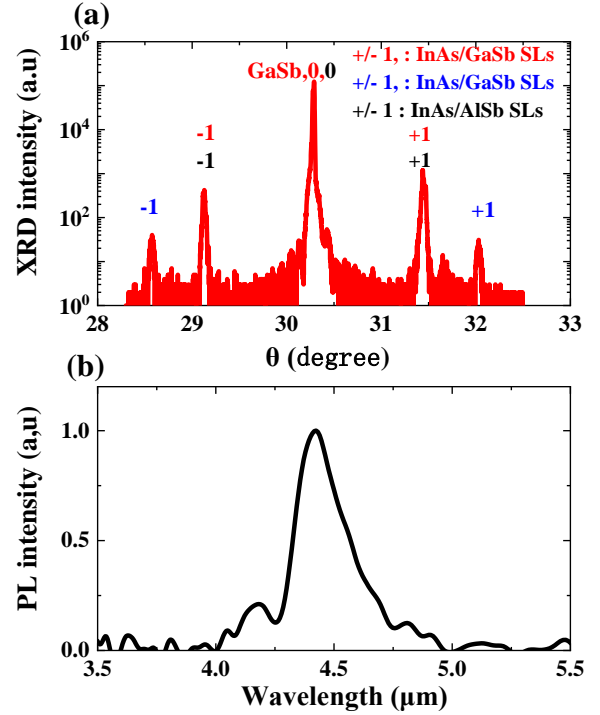

Fig. 2. (a)High resolution XRD scan of the epi-structure. (b) Normalized PL spectra measurement of the sample at 77K.

Fig. 2(a) shows the high resolution X-ray diffraction (HRXRD) scan curves of the grown sample. A series of sharp satellite peaks were identified that indicate good crystal quality, the 0th order satellite peaks for InAs/GaSb and InAs/AlSb SLs are very close to GaSb substrate which indicated good lattice-match, and the ±1th order satellite peaks of InAs/GaSb absorber SL and InAs/AlSb SLs are also very close (marked with red and black numbers). In addition, the ±1th order satellite peaks of bottom contact InAs/GaSb SL were also identified (marked with blue numbers). The periodic thickness of these SLs was calculated using [25]: d=λ/(2Δθcosθ), where λ is the wavelength of the X-ray, Δθ is the interval between two adjacent satellite peaks, and θ is the Bragg angle of the superlattice. From the calculated results, we found that the single periodic thickness of InAs/GaSb absorber SL and InAs/AlSb SL are both about 4.42 nm, and for InAs/GaSb bottom contact SL the periodic thickness is around 2.96 nm, which are very close to the designed thickness in Fig. 1(a). The photoluminescence (PL) of the sample was measured at 77 K as shown in Fig. 2(b). Here, a 532 nm laser was used as excitation source and a step-scan Fourier transform infrared spectrometer (FTIR) equipped with MCT detector was used to detect the emission spectra. An emission peak around 4.4 μm was observed, which corresponds to the transition between ground states in the absorption layer. The PL spectra presented a narrow full-width at half-maximum (FWHM) of 16 meV, which is also an indication of good crystalline quality.

After the material growth, the wafer was processed into a set of different sizes of mesa-shaped devices. Standard photolithography and citric-based wet chemical etchant ($C_6H_8O_7$:$H_3PO_4$:$H_2O_2$:$H_2O$=1:1:4:16) were used to define the mesa, and the etch was stopped at 100nm thick InAs/GaSb bottom contact layer. Ti/Pt/Au (20 nm/20 nm/80 nm) metal layers were deposited for both top and bottom contact layer to form ohmic contact by using e-beam evaporation, and mesa surfaces were passivated with SU-8. After that, a ground source–ground (GSG) coplanar waveguide (CPW) pad with an air-bridge structure was electroplated on 2 μm thickness SU-8 for high frequency measurement. Fig. 1(c) shows the schematic diagram of a fabricated device.

## III. ELECTRICAL AND OPTICAL CHARACTERIZATIONS

The temperature-dependent dark current density-bias voltage (J-V) characteristics of a 40 μm diameter device were measured and shown in Fig. 3(a). As temperature increases from 77K to 300K, the dark current density varies from $1.44 \times 10^{-5}$ A/cm$^2$ to

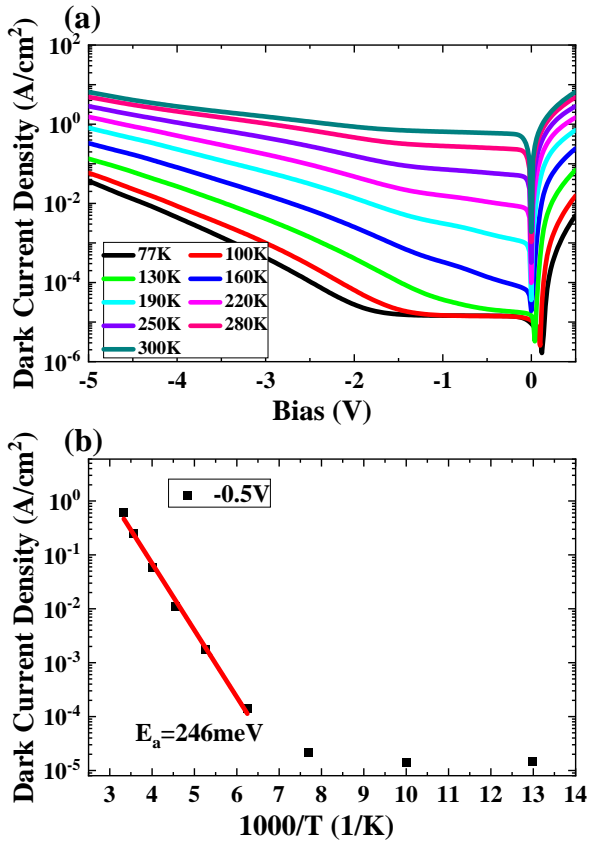

Fig. 3. (a) Dark current density as a function of temperature of a 40 μm diameter device. (b) Arrhenius plot of dark current density under -0.5V.

0.6 A/cm$^2$ under -0.5V. Fig. 3(b) shows an Arrhenius plot of temperature-dependent current density under -0.5V. The linear fit at high temperature region (160 K-300 K) yields an activation energy of 246 meV, which is close to the effective bandgap (about 275 meV to 221 meV from 77K to 300K) of

InAs/GaSb SL absorption layer, indicating that dark current is dominated by a diffusion component. In the low temperature region (77 K-130 K), the dark current becomes temperature-insensitive, suggesting that dark current at low bias could be mainly contributed by surface leakage and background radiation in the low temperature station.

The responsivity spectra of the device was measured by a FTIR in top-illuminated configuration and calibrated by a blackbody source. Fig. 4(a) shows the responsivity of the device under zero bias from 77 K to 300 K. The fluctuations around 2.7 μm, 3.4 μm and 4.2 μm are due to water vapor and carbon dioxide gas absorption in the atmosphere. The cutoff wavelength of device at 77 K is about 4.5 μm, which is consistent to the peak wavelength of the PL result shown in Fig. 2(b) and the peak responsivity is about 0.5 A/W. As the temperature increases to 300K, the cutoff wavelength shifts to about 5.6 μm and the peak responsivity reduces to 0.10 A/W. From 77 K to 190 K, the responsivity increases with temperature, this can be attributed to the shrinkage of the InAs/GaSb SL effective bandgap, which causes a larger absorption. However, for temperature above 190K, the responsivity drops rapidly with increasing temperature, which results from the enhanced carrier recombination rate and lower carrier lifetime at high temperature. The specific detectivity $D^*$ of the device was calculated by:

$$D^* = R_\lambda \sqrt{A} (2qI + \frac{4k_BT}{R})^{-1/2} \quad (1)$$

where q is the electronic charge, $k_B$ is Boltzmann's constant, T is the temperature of the device, R is the resistance under the bias, $R_\lambda$ is the responsivity, I is dark current and A is the area of device. Fig. 4(b) shows $D^*$ of the device under zero bias from 77K to 300K, a peak $D^*$ of $2.5 \times 10^{11}$ cm · $Hz^{1/2}$/W was achieved at 77 K. As temperature increases to 300K, the peak $D^*$ decreases to about $3 \times 10^8$ cm · $Hz^{1/2}$/W.

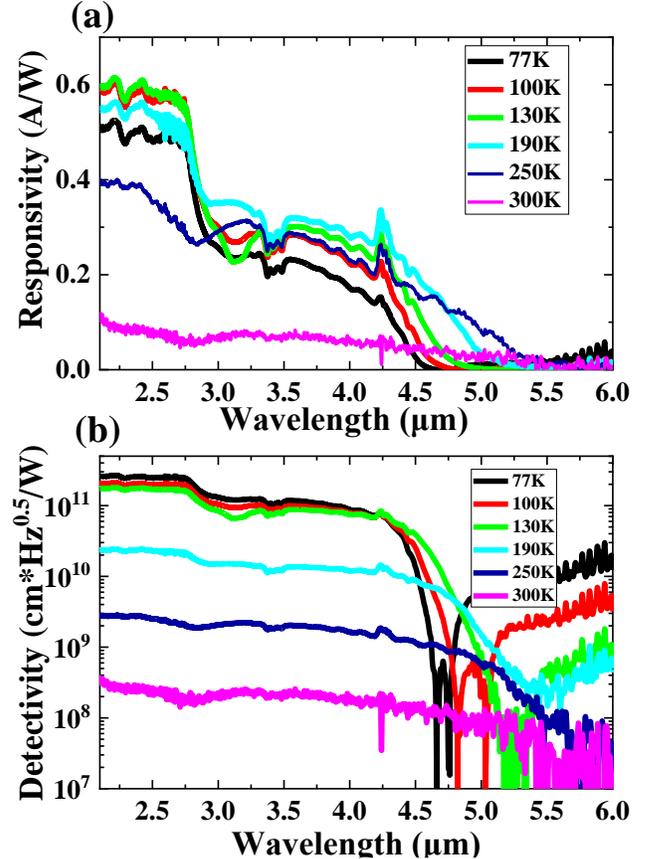

Fig. 4. (a) Responsivity and b) specific detectivity of device under 0V from 77K to 300K.

## IV. HIGH SPEED PERFORMANCE

For high speed characterization of the devices, a calibrated lightwave component analyzer (LCA) was used to test the bandwidth of the devices. Intensity-modulated light at 1550 nm generated by LCA was focused by a lensed fiber and coupled into the devices under test. The photoresponse of devices was collected by a Ground-Source-Ground (GSG) probe. The RF and DC components of the electrical signal were separated by an RF bias tee, the RF signal was returned to LCA, and the DC part was connected to a source meter which provides DC bias and measures photocurrent. It is noted that there may be some difference in term of 3dB bandwidth by characterizing with 1550nm light source instead of mid infrared light source. However, it is expected that 1550nm light source should be fully absorbed in the InAs/GaSb, and no additional photo-generated carrier would be created in InAs/AlSb drift layer, thus the absorption profile should be very similar to UTC PDs.

Fig. 5(a) shows the frequency response of a 20 μm diameter device under different biases at 300 K. A bias-dependent response is observed, the 3dB bandwidth increase rapidly from

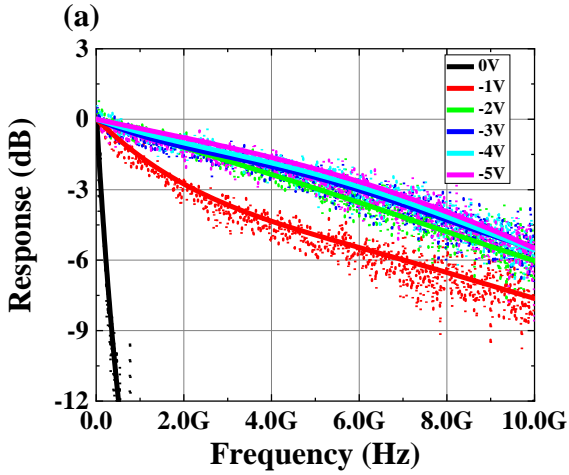

(a)

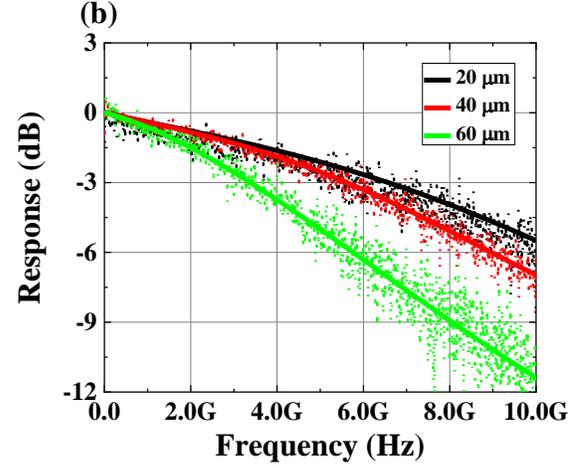

(b)

Fig. 5. (a) Frequency response of a 20 μm diameter device at various biases under 300K. (b) Frequency response of devices with different sizes under at -5V (dot line is the measured data and the solid line is polynomial fitting curve).

TABLE I
EXTRACTED PARAMETERS OF DEVICE FROM EQUIVALENT CIRCUIT MODEL

| Device diameter | Cj(fF) | Rp(Ohm) | Rs(Ohm) |
|---|---|---|---|
| 20 | 70 | 6500 | 32.5 |
| 40 | 240 | 5000 | 14 |
| 60 | 500 | 3200 | 8.5 |

120 MHz to 5.1 GHz when the bias increases from 0V to -2V as the InAs/AlSb drift layer is fully depleted. With bias further increases to -5V, the 3dB bandwidth reaches to 6.58 GHz. The slightly increasing 3 dB bandwidth with reverse bias beyond -2V is due to faster electron transport in the drift layer with increased reverse bias. By applying a higher reverse bias, the 3dB bandwidth could further increase, however, the increasing dark current will degrade the signal to noise ratio of the device and also breaks down the device. Fig. 5(b) shows the frequency response of three devices with different sizes under -5V at 300K. The measured 3dB bandwidth of devices with diameter of 20 μm, 40 μm and 60 μm are 6.58 GHz, 5.62 GHz and 3.39 GHz, respectively. The weak diameter-dependent bandwidth indicates that the 3dB bandwidth of these devices should be dominated by transit time. To further investigate the bandwidth limit mechanism, we also measured $S_{11}$ parameter of these devices. An equivalent circuit model shown in Fig. 6(a) was used to fit measured $S_{11}$ parameter, where $C_j$ is the junction capacitance, $R_p$ is junction resistance, $R_s$ is the series resistance, $I_p$ is AC current source, $R_L$ is load resistance. Fig. 6(b) – Fig. 6(d) show the measured and fitting $S_{11}$ curves of these three devices. With these extracted circuit model parameters, a theoretical RC time limited frequency response is plot in Fig. (6e). The RC limited 3dB bandwidth of 20 μm, 40 μm and 60 μm devices are calculated to be 27.85 GHz,10.47 GHz and 5.53 GHz, respectively. These RC limited bandwidth values are much larger than the corresponding measured bandwidth values for each size of devices, which indicates that the 3dB bandwidth of these devices is limited by carrier transit time. The extracted circuit model parameters are summarized in Table I.

To further investigate the transit process of the carriers in the device, the frequency response of a 20 μm diameter device at different temperatures and different biases was measured. Fig. 7(a) shows the measured frequency response under -5 V from 77 K to 300 K, and the measured 3dB bandwidth is summarized in Fig. 7(b). As seen in Fig. 7(b), the 3dB bandwidth versus temperature under different biases behaves differently. For the device under low bias, such as -2V, the 3dB bandwidth increases significantly at high temperature. Under high reverse bias (such as -5 V), 3dB bandwidth doesn't change significantly as temperature increases. That is mainly due to the unexpected conduction band discontinuity at the interface between absorption layer and drift layer. Under low bias condition, the unexpected barrier in conduction band could block the electrons transport. Thermal emission at higher temperature will help electrons to overcome the barrier. Under high bias condition, the barrier will not block the carrier transport, given the high electrical field in depletion region. The 3dB bandwidth is not sensitive to temperature in that case.

Based on the analysis above, the 3dB bandwidth of the device at lower voltages is limited by carrier trapping, while it is limited by electron transit time including the diffusion time in absorption layer and the drift time in drift layer at higher voltages. To further improve the speed of MWIR UTC PD, the InAs/AlSb drift layer should be further optimized to reduce the expected barrier between absorption layer and drift layer.

Table II summarized a performance comparison between high speed MWIR photodiodes with different structures or materials. It should note that some parameters in others works are not reported directly, and are extracted from the figures shown in the references. Comparing with other research, the 3dB bandwidth of our device shows much higher performance. However, there are still remaining problems associated with the relatively small responsivity and slightly larger dark current for the UTC MWIR PDs as compared to interband cascade infrared photodetectors [12]. Engineering the design of drift layers could enable the device working at low bias or even zero bias. Using resonant cavity structure or waveguide structure could improve the responsivity of the device [26]. Moreover, the bandwidth can be further improved by using highly resistive or semi-insulated substrate, reducing the device size, and applying waveguide structures to better balance between the bandwidth and the responsivity.

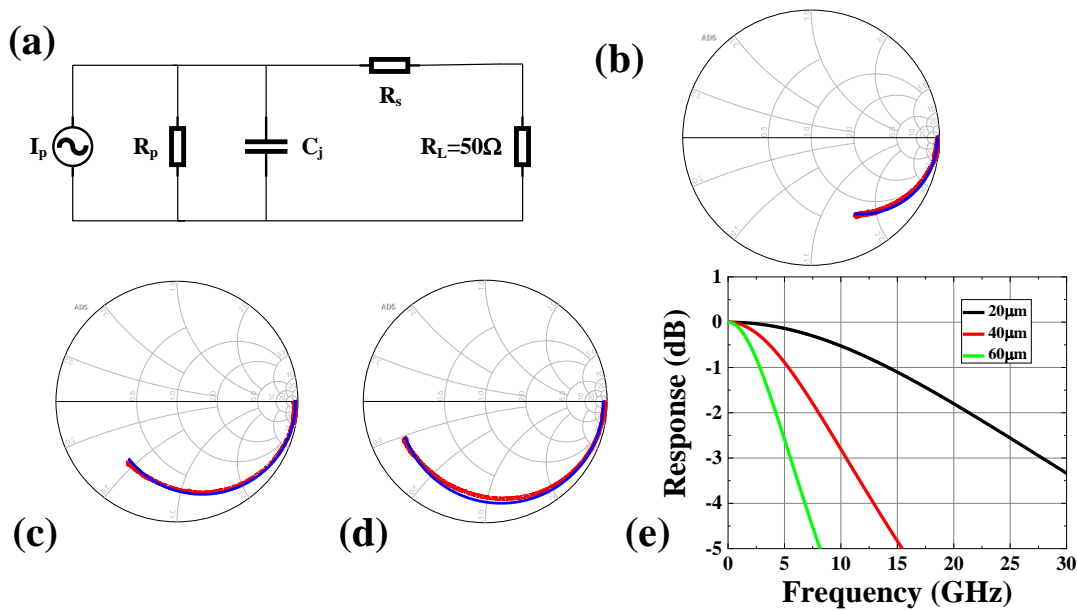

Fig. 6. (a) Equivalent circuit model of the UTC PD for $S_{11}$ parameter fitting. Measured (red line) and fitting (blue line) $S_{11}$ parameter from 10 MHz to 30 GHz of (b) 20 μm, (c) 40 μm and (d) 60 μm diameter device under -5V. (e) Calculated RC limit frequency response with extracted circuit model parameters.

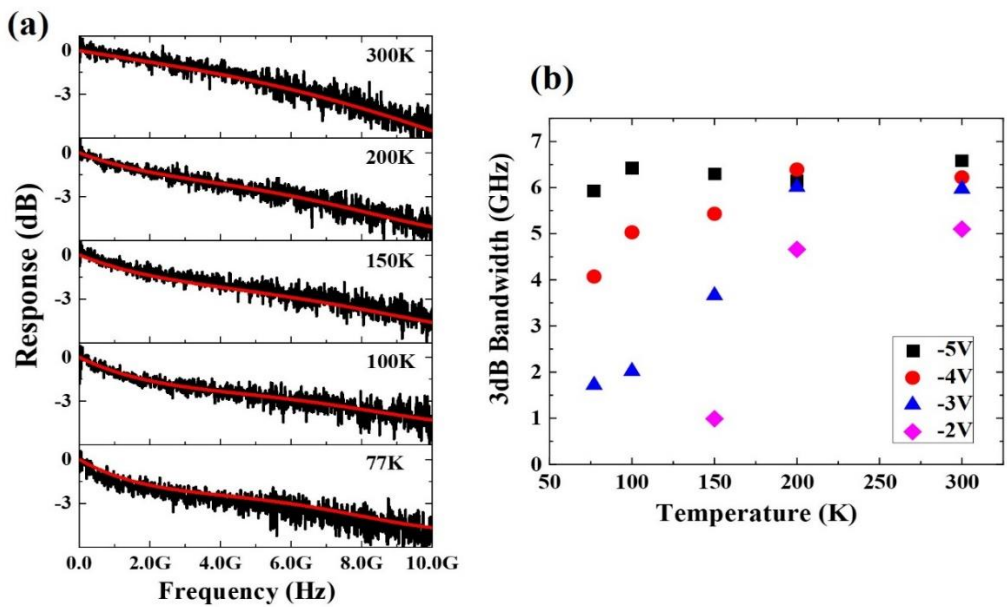

Fig. 7. (a) Measured frequency response of a 20 µm diameter device under -5V at different temperature. (b) 3 dB bandwidth of a 20 µm diameter device varies temperature under different biases.

TABLE II
PERFORMANCE COMPARISON OF HIGH SPEED MWIR PHOTODETECTORS, ALL DATA SHOWN ARE AT 300K OR AT ROOM TEMPERATURE

| Device parameters | Cutoff wavelength [µm] | DC density [A/cm²] | Peak responsivity [A/W] | Peak D* [Jones] | 3dB bandwidth [GHz] |
|---|---|---|---|---|---|
| ICIP [12] | 4.2 | 0.2(1 V) | 0.3 (0V) | $1.5 \times 10^9$ (0V) | 1.3 |
| ICIP [13] | 5.3 | 7 (-0.5 V) | 0.1 (0V) | N/A | 2.4 |
| QWIP [16] | 4.9 | NA | 0.1(-5V) | $1 \times 10^7$ (-5V) | 5 |
| QWIP [17] | 9 | 973.68 (-0.7V) | 0.08 (0.7V) | $3.3 \times 10^7$(0.7) | 5 |
| This work | 5.6 | 0.6 (-0.5V) | 0.1 (0V) | $3 \times 10^8$(0V) | 6.58 |

## V. CONCLUSION

In summary, an InAs/GaSb T2SL based MWIR UTC PD for high speed application has been demonstrated in this work. The cutoff wavelength of around 5.6 µm and the 3dB bandwidth of 6.58 GHz are obtained for a 20 µm diameter device at 300K. The equivalent circuit analysis shows the 3dB bandwidth of device is limited by transit time of optically generated carrier. With further material and device structure optimization, one can envision the prospects of using these UTC PDs in future free space optical communication, frequency comb spectroscopy and some other emerging areas requiring high speed MWIR receivers.

**Jian Huang** received the B.S. degree in Materials science and Engineering from Xidian University in 2017. He is currently Ph.D student in the school of information science and technology at ShanghaiTech University. His research interests focus on Mid infrared photodetector.

**ZhiYang Xie** and received the B.S. degree in Optoelectronic information science and Engineering from Huazhong University of Science and Technology (HUST) in 2017. In 2017, he joined School of Information Science and Technology, ShanghaiTech University, China. He is currently working towards his Ph.D degree in Electronics Science and Technology at ShanghaiTech University. His research interest focus on high speed photodiodes and microwave photonics.

**Yaojiang Chen** received the B.S. degree in electrical engineering from Shanghai Jiao Tong University (SJTU) in 2015. He is currently working towards his Ph.D degree in microelectronics and solid-state electronics at ShanghaiTech University. His research interests focus on high speed photodiodes and applications in analog optical links.

**John E. Bowers** received the M.S. and Ph.D. degrees from Stanford University. He was with the AT&T Bell Laboratories. He is currently the Director of the Institute for Energy Efficiency and a Professor with the Departments of Electrical and Computer Engineering and Materials, University of California at Santa Barbara, Santa Barbara. His research interests are primarily concerned with silicon photonics, optoelectronic devices, optical switching and transparent optical networks, and quantum dot lasers. He is also a member of the National Academy of Engineering and the National Academy of Inventors. He is also a fellow of the OSA and the American Physical Society. He was a recipient of the IEEE Photonics Award, the OSA/IEEE Tyndall Award, the IEEE LEOS William Streifer Award, and the South Coast Business and Technology Entrepreneur of the Year Award

**Baile Chen** received the bachelor's degree in physics from the Department of Modern Physics, University of Science and Technology of China, Hefei, China, in 2007, and the master's degree in physics and the Ph.D. degree in electrical engineering from the University of Virginia, Charlottesville, VA, USA, in 2009 and 2013, respectively. In 2013, he joined Qorvo Inc., Hillsboro, OR, USA, as an RF Product Development Engineer, where he is involved in various RF power amplifiers and BAW filters for RF wireless communication systems. In 2016, he joined the School of Information Science and Technology, ShanghaiTech University, Shanghai, China, as a Tenure-Track Assistant Professor and a PI. His current research interests include III–V compound semiconductor materials and devices and silicon photonics.